\documentclass[aps,prb,twocolumn,groupedaddress,showpacs]{revtex4-1}

\usepackage{graphicx}
\usepackage{amsmath}
\usepackage{amsbsy}
\usepackage{dcolumn}
\usepackage{bm}
\usepackage{braket}
\usepackage[labelsep=period]{caption}
\usepackage{subcaption}

\begin{document}


\title{First-principles study on interface magnetic structure in Nd${}_2$Fe${}_{14}$B/(Fe, Co) exchange spring magnets} 



\author{Nobuyuki Umetsu}
 \affiliation{Department of Applied Physics, Tohoku University, Aoba 6-6-05, Aoba-ku, Sendai 980-8579, Japan}
\author{Yuta Toga}
 \affiliation{ESICMM, National Institute for Materials Science, Sengen 1-2-1, Tsukuba 305-0047, Japan}
\author{Akimasa Sakuma}
 \affiliation{Department of Applied Physics, Tohoku University, Aoba 6-6-05, Aoba-ku, Sendai 980-8579, Japan}
 


\date{\today}

\begin{abstract}
The magnetic properties of Nd${}_2$Fe${}_{14}$B (NFB)/transition metal (TM = Fe, Co) multilayer systems are studied
on the basis of first-principles density functional calculations.
Assuming a collinear spin structure, we optimize the model structure under a variety of crystallographic alignments of the NFB layer,
and analyze the mechanism of interface magnetic coupling.
Improvements in remanent magnetization compared to that of single  NFB
are observed in NFB(001)/Fe, NFB(110)/Fe, and NFB(100)/Co.
On the other hand, in NFB(100)/Fe,
remanence degradation due to the anti-parallel magnetization alignment between NFB and Fe layers is observed. 
In this system, which has the shortest optimized interlayer distance among all considered systems,
an itinerant electron magnetism is required around the interface to lower the total energy, and accordingly,
anti-ferromagnetic coupling is preferred.
The significant difference in property between NFB(100)/Fe and NFB(100)/Co is attributed to
the difference between their interface structures, optimized interlayer distances, and magnetic stiffness of TM layers.
\end{abstract}

\pacs{}

\maketitle 

\section{Introduction}

Much effort has been made to improve the properties of permanent magnets,
whose figure of merit is the maximum energy product $(BH)_{\rm max}$ that increases with coercivity and remanent magnetization.
Exchange spring magnets \cite{CoehoornMooijWaard1989,KnellerHawig1991},
nanocomposite materials consisting of hard and soft magnetic phases coupled by exchange interaction,
have been promising as high-performance magnets \cite{SkomskiCoey1993},
and Nd--Fe--B-based exchange spring magnets are particularly attractive from the viewpoint 
of  low rare-earth metal content.
In theoretical studies, 
$(BH)_{\rm max}$ values of 0.6$\sim$1.0 MJm${}^{-3}$ for Nd--Fe--B-based magnets  have been predicted \cite{LeineweberKronmueller1997,Saiden2014,Zhao2006,Li2015},
but such high energy-product values have been difficult to achieve in real materials \cite{Shindo,Wang2002,Liu,Lupu2009,
Su2011,Urse2011,OgawaKoikeMizukamiOoganeAndoMiyazakiKato2012,Jun2013,CuiTakahashiHono}.
To obtain greater values of $(BH)_{\rm max}$ in real materials, the advanced design of nanostructures is required.
Although the optimal grain size and multilayer thickness of exchange spring magnets have been intensively studied \cite{SchreflFidlerKronmdotuller1994,Sabiryanov1998, AmatoPiniRettori1999,Asti2004,Fullerton,Jiang2014,Horikawa2014,Pernechele2014},
little is known about other crucial factors affecting the magnetic properties of exchange spring magnets.

Based on a first-principles study, Toga et al. pointed out that the magnetic properties of Nd${}_2$Fe${}_{14}$B (NFB)/bcc-Fe multilayer systems
strongly depend on the crystallographic alignments of NFB and Fe layers \cite{TogaMoriyaTsuchiuraSakuma2011}.
It was shown that the NFB layer of the (001) plane is ferromagnetically coupled with Fe layers,
but the (100) plane is anti-ferromagnetically coupled.
Recently, these predictions were confirmed by Ogawa et al. \cite{Ogawa2015} by performing ferromagnetic resonance measurements. 
Their results imply that random crystallographic alignments possibly deteriorate  magnetic performances because of low remanent magnetization.
However, the reason for the drastically different results depending on crystallographic alignments is not known at present.
Moreover, it has been recognized that structure optimizations performed in a previous study \cite{TogaMoriyaTsuchiuraSakuma2011} are not sufficient,
because those structures are only optimized with respect to the interlayer distance between the NFB and Fe layers,
but not with respect to cell volume, cell shape, and ion sites.
In order to identify and present reliable guidelines for fabricating high-performance magnets,
the adequate optimization of nanostructures is desirable.

In the present work, we theoretically study the interface magnetic structure in exchange spring magnets.
Our aim is to optimize the structure of the NFB/transition metal [TM = (bcc-)Fe, (hcp-)Co] multilayer systems
for a variety  of crystallographic alignments of NFB layers,
and to analyze those interface electronic structures in order to understand crucial factors that determine the properties of high-performance magnets.
In particular, we focus on NFB(100)/TM, which shows significantly different properties between TM = Fe and TM = Co systems. 

In our calculations, we assume a collinear spin structure between local moments at NFB layer and TM layer.  Actually, there may be a possibility of non-collinear structure, but credible results for non-collinear structures are hard to obtain in such a case of finite sized super-cell model with periodic boundary condition.  Therefore, a detailed spin structure is not predicted from our model, but the results suggest a necessity to reconsider the tacit postulation that the interface coupling in exchange spring magnets is always ferromagnetic.

This paper is organized as follows.
In Section II, we present our model and explain the method of structure optimization.
In Section III, the results of our calculation are presented and discussed.
Finally, the summary and conclusions of our work are described in Section IV.

\section{Model and Method}

\begin{figure}[htbp]
	\begin{minipage}[b]{0.32\columnwidth}
		\centering
		\subcaption{\footnotesize{NFB(001)/Fe(100)}}
		\includegraphics[width=1.0\columnwidth]{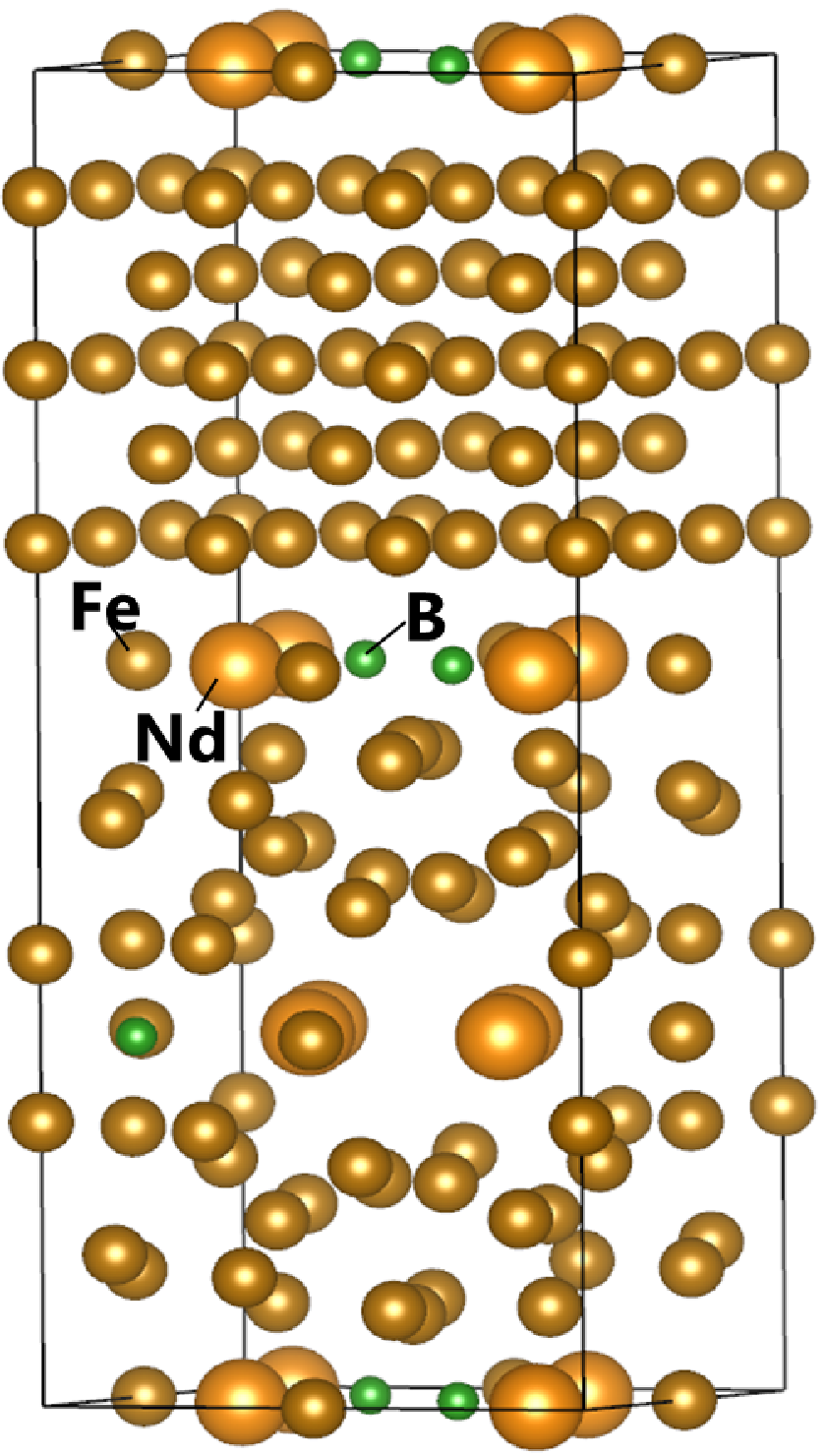}
	\end{minipage}
	\begin{minipage}[b]{0.32\columnwidth}
		\centering
		\subcaption{\footnotesize{NFB(100)/Fe(110)}}
		\includegraphics[width=1.0\columnwidth]{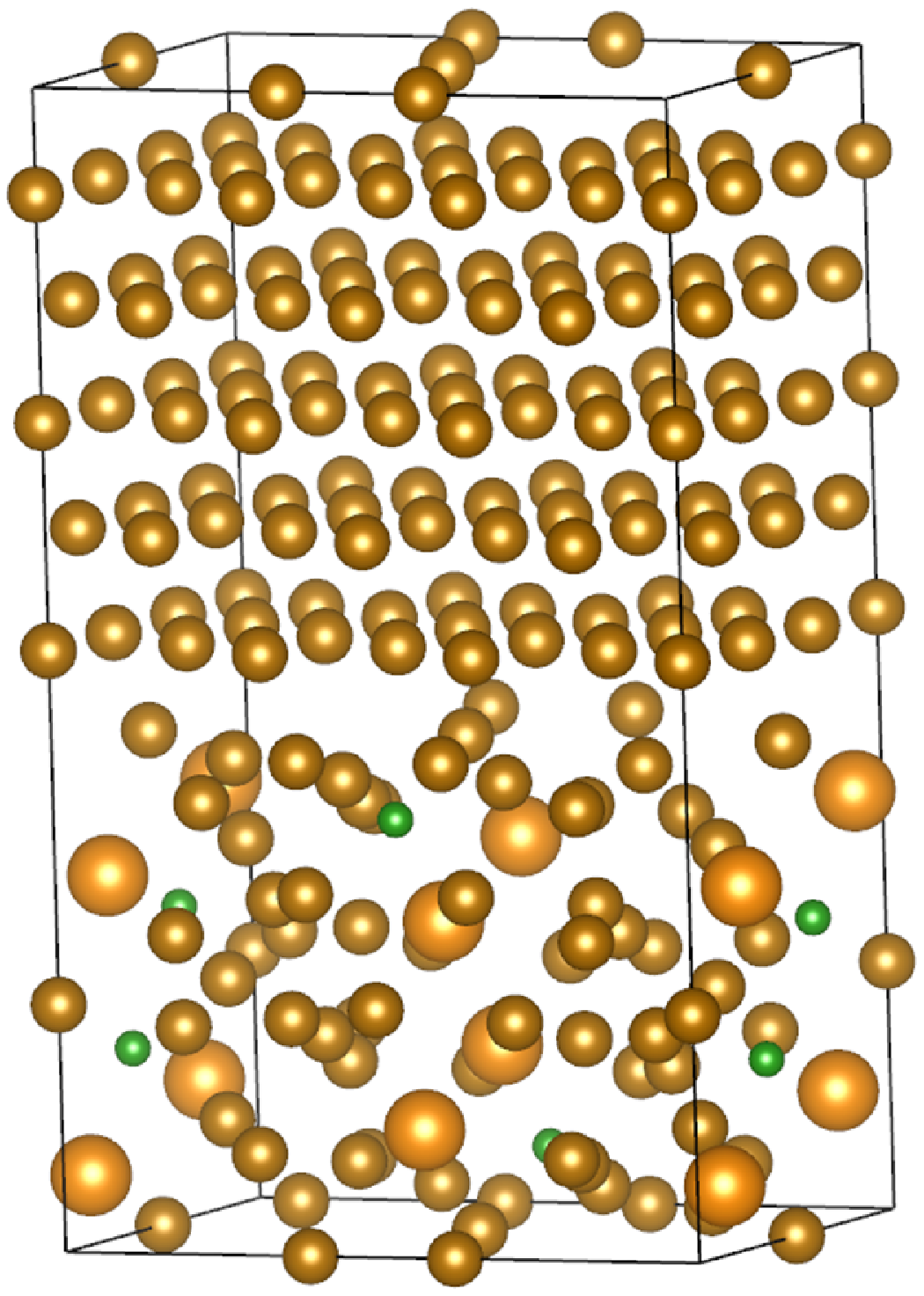}
	\end{minipage}
	\begin{minipage}[b]{0.32\columnwidth}
		\centering
		\subcaption{\footnotesize{NFB(110)/Fe(112)}}
		\includegraphics[width=1.0\columnwidth]{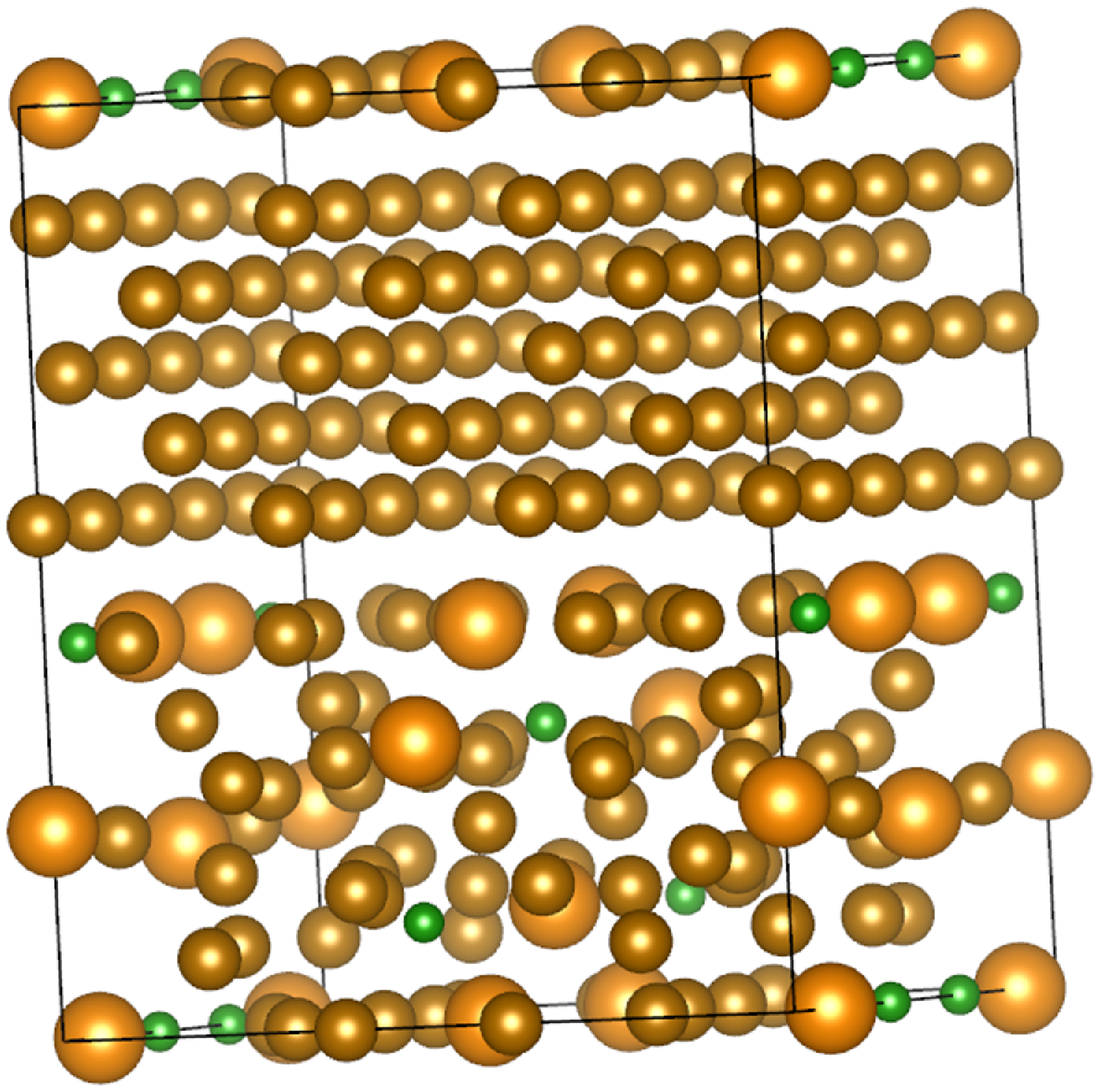}
	\end{minipage}
	\begin{minipage}[b]{0.32\columnwidth}
		\centering
		\subcaption{\footnotesize{NFB(001)/Co(0001)}}
		\includegraphics[width=1.0\columnwidth]{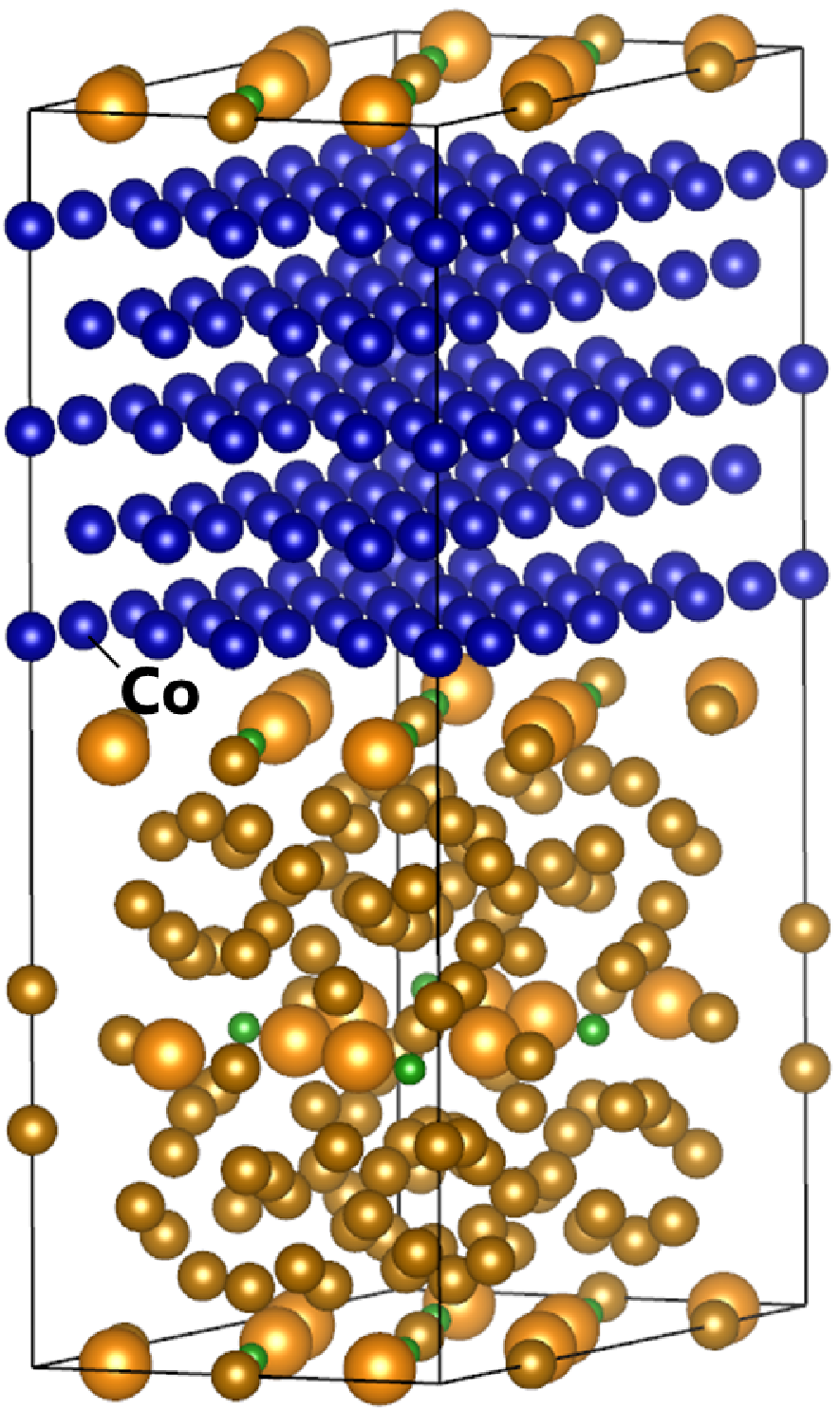}
	\end{minipage}
	\begin{minipage}[b]{0.32\columnwidth}
		\centering
		\subcaption{\footnotesize{NFB(100)/Co(0001)}}
		\includegraphics[width=1.0\columnwidth]{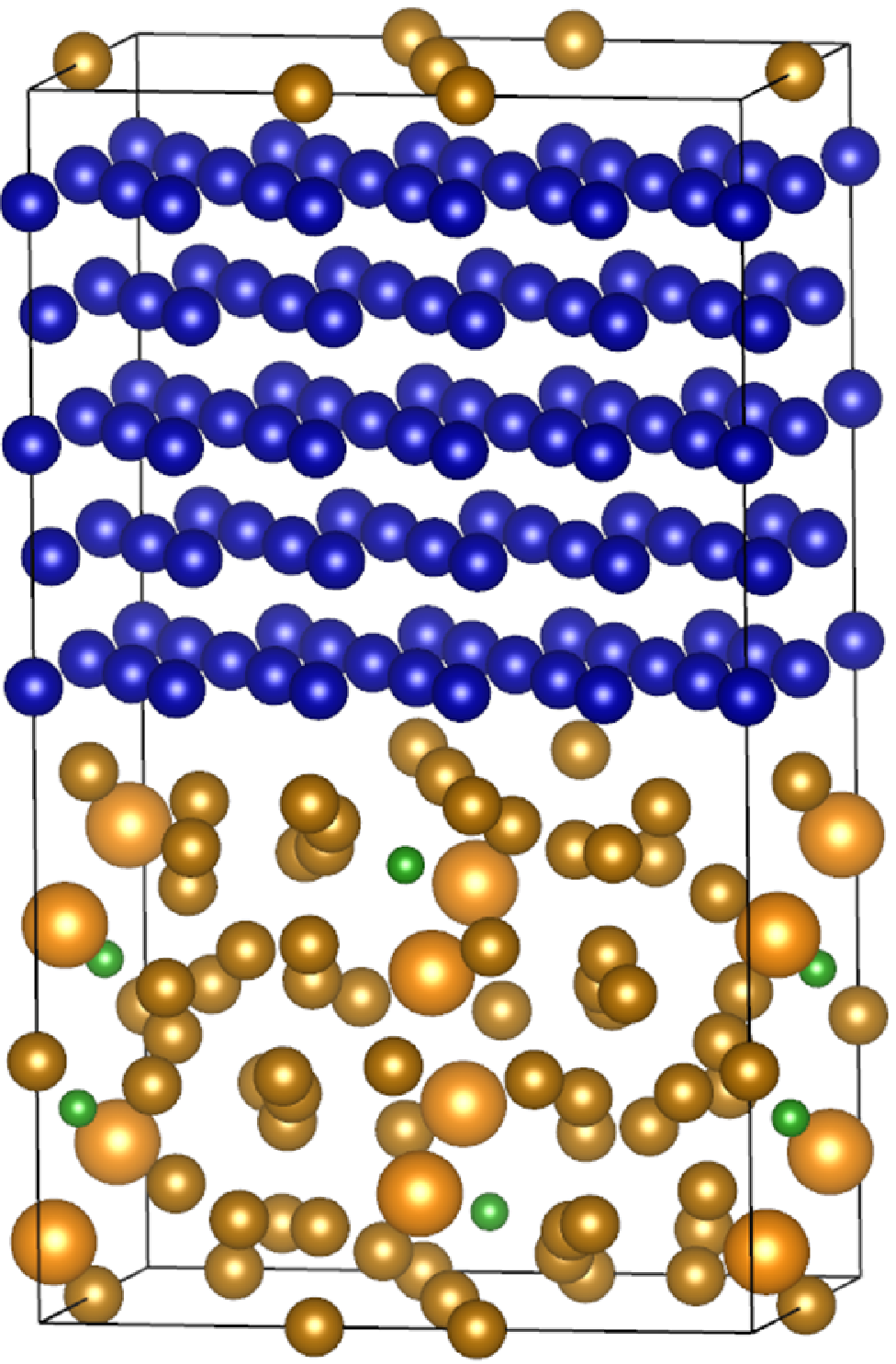}
	\end{minipage}
	\begin{minipage}[b]{0.32\columnwidth}
		\centering
		\subcaption{\footnotesize{NFB(110)/Co(0001)}}
		\includegraphics[width=1.0\columnwidth]{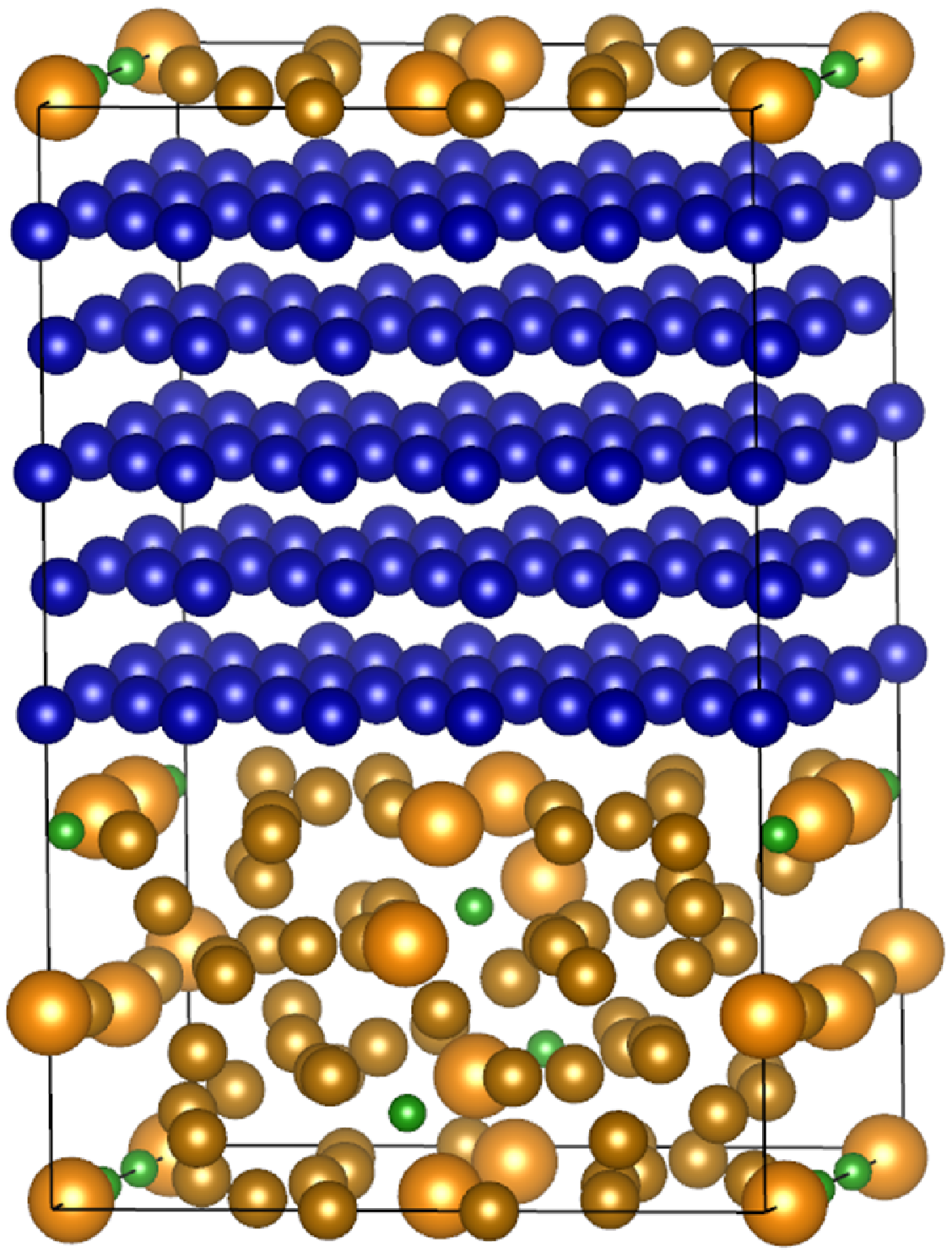}
	\end{minipage}
	\caption{\raggedright
	Composite system models.
	Two unit cells of NFB are incorporated in the supercell of NFB(001)/Co(0001),
	while one unit cell is incorporated in the others.
	These figures are plotted using VESTA \cite{MommaIzumi2008}.}
	\label{fig:model}
\end{figure}

\begin{figure}[htbp]
	\begin{minipage}[b]{0.48\columnwidth}
		\centering
		\subcaption{bcc-Fe}
		\includegraphics[width=0.95\columnwidth]{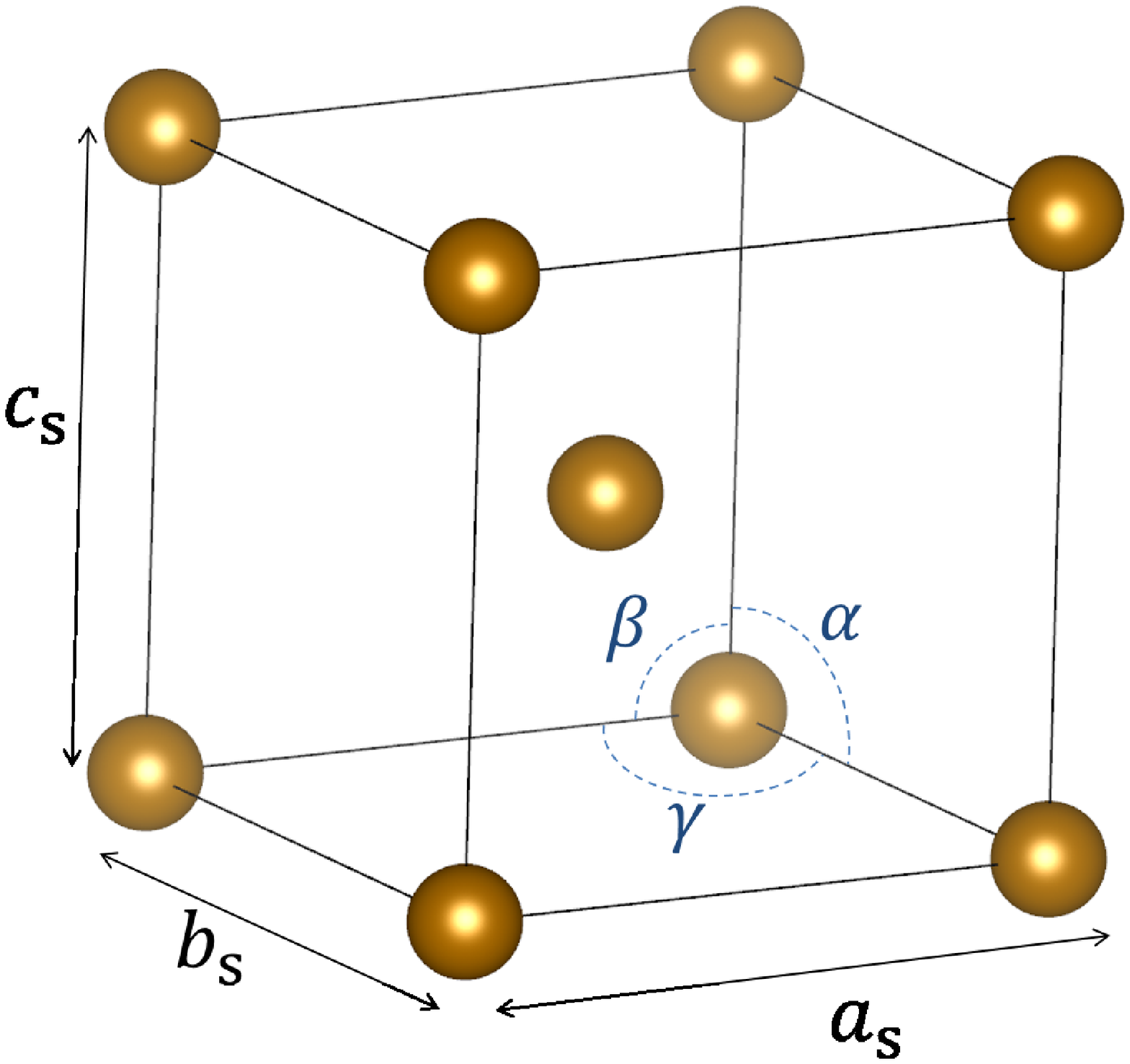}
		\label{fig:001fe}
	\end{minipage}
	\begin{minipage}[b]{0.48\columnwidth}
		\centering
		\subcaption{hcp-Co}
		\includegraphics[width=0.95\columnwidth]{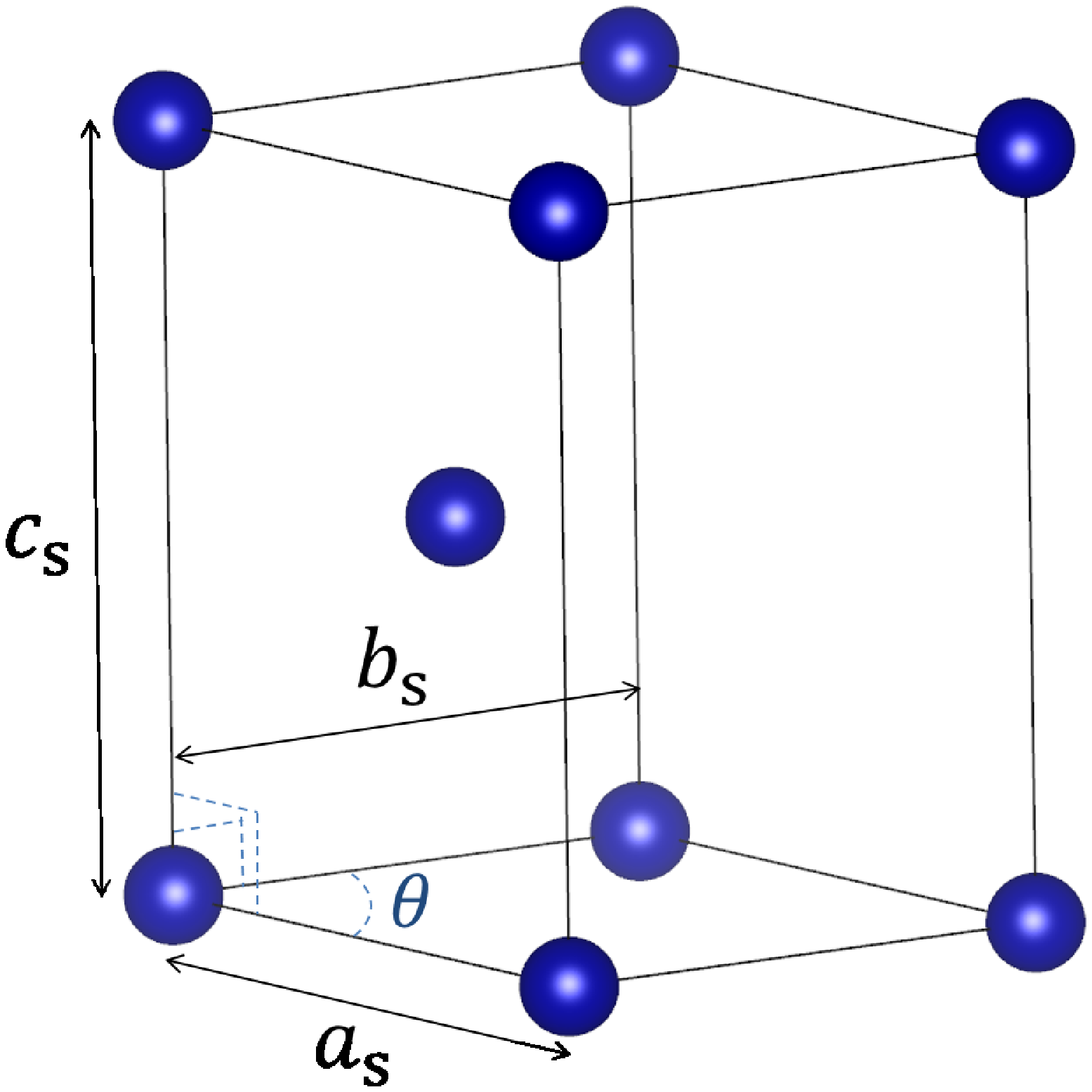}
		\label{fig:100fe}
	\end{minipage}
	\caption{
	Definitions of lattice parameters (see TABLE \ref{table:parameter}).
\label{fig:feco}}
\end{figure}

We study the electronic structure of NFB/TM multilayer systems, as shown in FIG. \ref{fig:model}.
The lattice constants of NFB are set to $a_{\rm h}=b_{\rm h}=8.8$ \AA \, and $c_{\rm h}=12.19$ \AA,
and ion sites are determined as per a previous study \cite{Herbst1991}.
The bcc-Fe and hcp-Co are assumed as structurally soft phases,
and these plane indices facing the interface are chosen so as to match the surface size of an NFB unit cell.
It is experimentally confirmed that the (001) plane and (100) plane of NFB layers
match the (100) plane and (110) plane of bcc-Fe layers, respectively,
but the other systems studied in this paper have not been well investigated experimentally.
The deformed lattice constants of these soft phases are shown in TABLE \ref{table:parameter}
(the definitions of parameters are shown in FIG. \ref{fig:feco}).
The thickness of the soft layer of our models corresponds to five atomic layers, which is different from that of previously reported models\cite{TogaMoriyaTsuchiuraSakuma2011}. 

\begin{table}[htbp]
\centering
\caption{\raggedright
Lattice constants of the soft magnetic phases of NFB/TM (TM = bcc-Fe, hcp-Co).
The definitions of parameters are shown in FIG. \ref{fig:feco}.
The experimental values for bcc-Fe are $a_{\rm s}=b_{\rm s}=c_{\rm s}=2.87$ \AA\, and $\alpha=\beta=\gamma=90^{\circ}$, and
those of hcp-Co are $a_{\rm s}=b_{\rm s}=2.51$ \AA, $c_{\rm s}=4.07$ \AA, and $\theta=60^{\circ}$.}
\label{table:parameter}
\begin{tabular}{c|c c c c c c c}
\hline 
\hline
 & $a_{{\rm s}}${[}\AA{]} & $b_{{\rm s}}${[}\AA{]} & $c_{{\rm s}}${[}\AA{]} & $\alpha${[}$^{\circ}${]} & $\beta${[}$^{\circ}${]} & $\gamma${[}$^{\circ}${]} & $\theta${[}$^{\circ}${]}\tabularnewline
\hline 
NFB(001)/Fe(100) & 2.93  & 2.93 & 2.87 & 90.0 & 90.0 & 90.0 & -\tabularnewline

NFB(100)/Fe(110) & 2.87  & 2.87  & 2.93 & 90.0 & 90.0 & 90.1 & -\tabularnewline

NFB(110)/Fe(112) & 2.87 & 2.87  & 2.88 & 90.0 & 90.0 & 90.2 & -\tabularnewline

NFB(001)/Co(0001) & 2.51 & 2.44 & 4.07 & - & - & - & 61.0 \tabularnewline

NFB(100)/Co(0001) & 2.51 & 2.51 & 4.07 & - & - & - & 60.2 \tabularnewline

NFB(110)/Co(0001) & 2.41 & 2.44 & 4.07 & - & - & - & 59.5 \tabularnewline
\hline
\hline
\end{tabular}
\end{table}

The electronic structures are determined with first-principles density functional calculations using a plane-wave basis set.
We use the Vienna Ab-initio Simulation Package (VASP) \cite{Kresse}.
The ionic potentials are described by the projector-augmented-wave (PAW) method and
the exchange-correlation energy of electrons is described within a generalized gradient approximation (GGA).
We use the exchange-correlation functional determined by Ceperly and Alder and parameterized by Perdew and Zunger,
with the interpolation formula according to Vosko et al. \cite{VoskoWilkNusair1980}.
The energy cutoff is 318.6 eV in all systems, and  
the Monkhorst-Pack $\bm{k}$-point meshes of $1\times 3 \times 3$, $3\times 3 \times 1$, and $3 \times 3 \times 3$
are selectively used depending on the model size.
Our computations are performed until the total energy change converges to below 10$^{-4}$ eV.
The collinear spin structures are assumed,
and magnetic stable states are determined by comparing the total energy for parallel magnetization alignment (PMA)
and anti-parallel magnetization alignment (APMA) between NFB and TM layers.
The distance between these layers is first optimized by using the fixed lattice constants shown in TABLE \ref{table:parameter}. 
By using force and stress, which are calculated using first principles,
we perform additional optimization with respect to the cell volume, cell shape, and ion sites.
For optimization with respect to ion sites,
we assume that only the ions facing the interface boundary are allowed to change positions.

\section{Results and Discussion}

\begin{figure}[htbp]
	\begin{minipage}[b]{0.494\columnwidth}
		\centering
		\includegraphics[width=1.0\columnwidth]{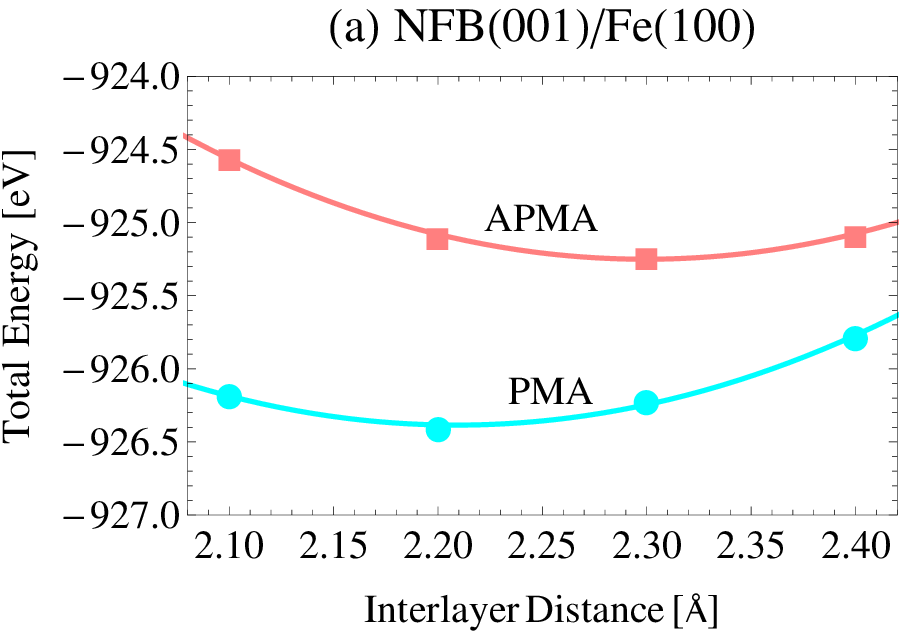}
	\end{minipage}
	\begin{minipage}[b]{0.494\columnwidth}
		\centering
		\includegraphics[width=1.0\columnwidth]{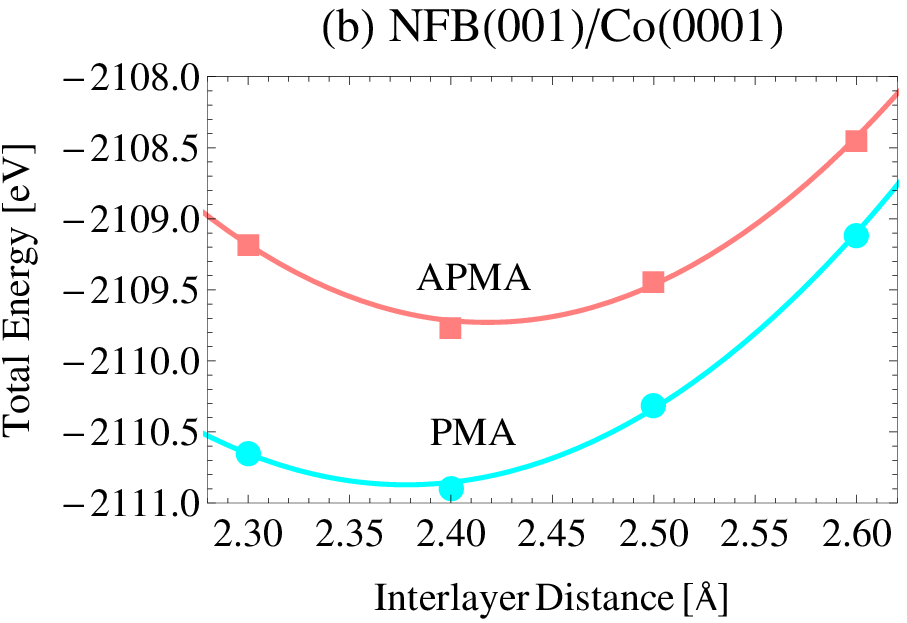}
	\end{minipage}
	\begin{minipage}[b]{0.494\columnwidth}
		\centering
		\includegraphics[width=1.0\columnwidth]{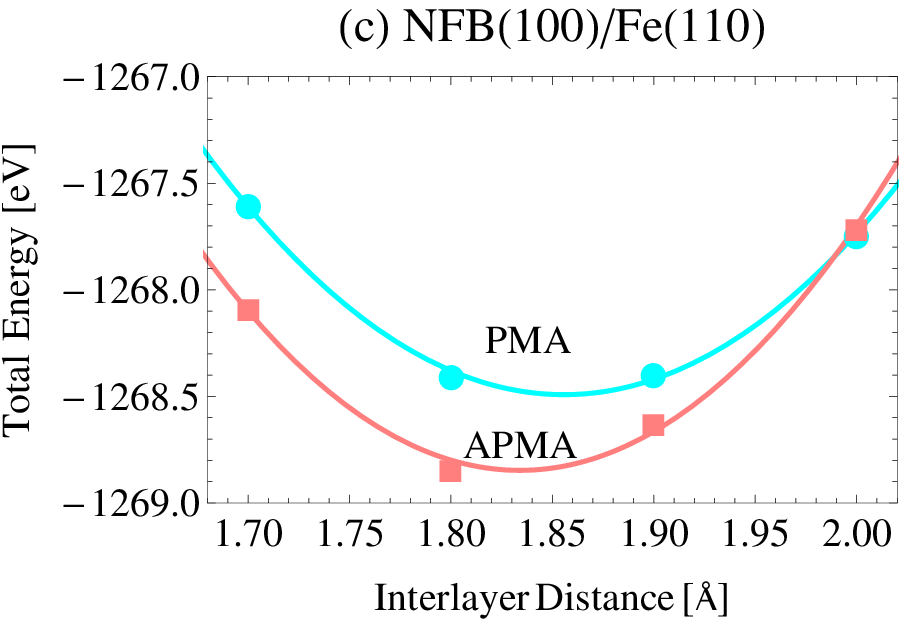}
	\end{minipage}
	\begin{minipage}[b]{0.494\columnwidth}
		\centering
		\includegraphics[width=1.0\columnwidth]{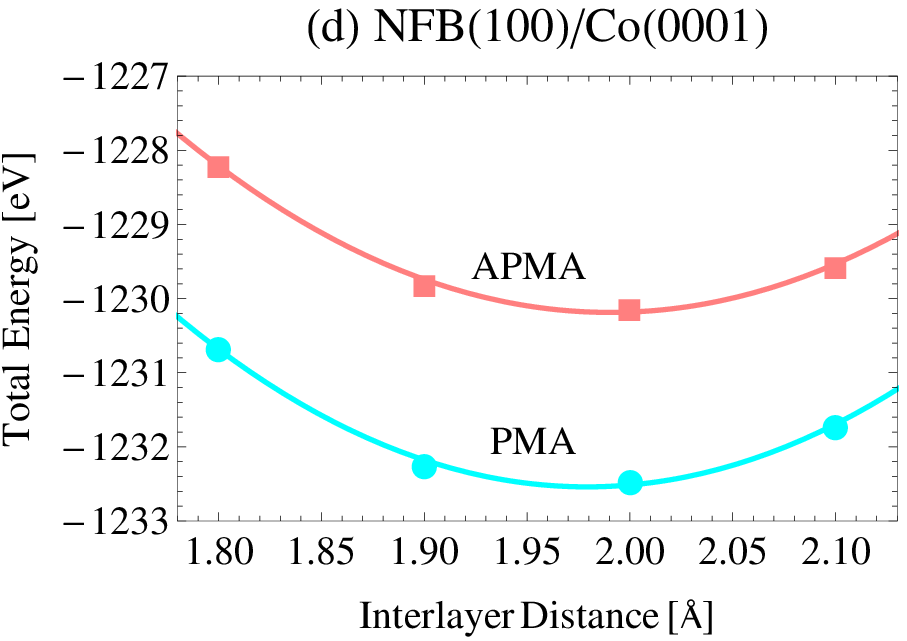}
	\end{minipage}
	\begin{minipage}[b]{0.494\columnwidth}
		\centering
		\includegraphics[width=1.0\columnwidth]{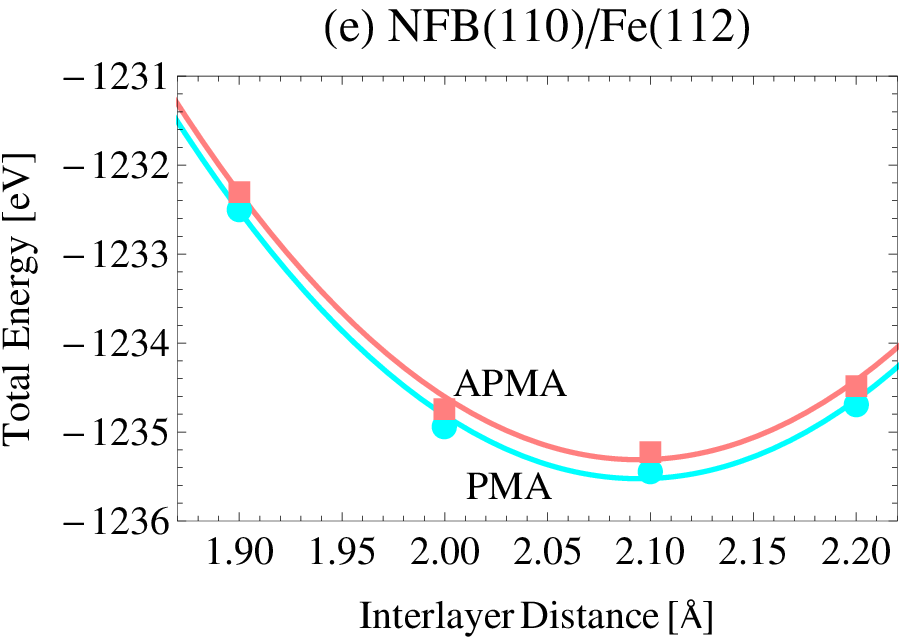}
	\end{minipage}
	\begin{minipage}[b]{0.494\columnwidth}
		\centering
		\includegraphics[width=1.0\columnwidth]{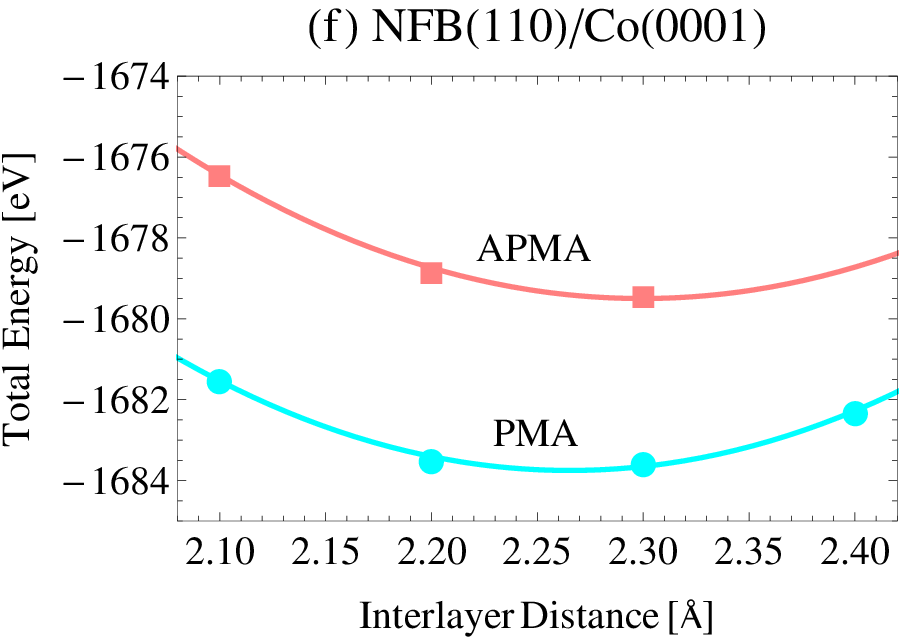}
	\end{minipage}
	\caption{\raggedright
	Total electronic energy of NFB/TM (results before optimization).
	The blue circles and red squares show the results for parallel magnetization alignment (PMA)
	and anti-parallel magnetization alignment (APMA), respectively.}
	\label{fig:freeng}
\end{figure}

In FIG. \ref{fig:freeng}, we show the interlayer distance dependence of the total electronic energy of NFB/TM for PMA and APMA.
These are the results before optimization with respect to the cell volume, cell shape, and ion sites,
while the magnetic properties (i.e., coupling constant and magnetization) after optimization considering these factors are summarized in TABLE \ref{table:opt}, in addition to the results before optimization.
The exchange coupling energy between magnetization $\bm{M}_1$ and $\bm{M}_2$ is written as
\begin{equation}
E_{\rm ex}=-J\bm{n}_1\cdot\bm{n}_2,
\end{equation}
where $J$ is coupling constant, and $\bm{n}_{1(2)}=\bm{M}_{1(2)}/|\bm{M}_{1(2)}|$ is an unit vector.
Since the difference of total energy for between APMA and PMA corresponds to the interface coupling energy,
coupling constant is obtained by
\begin{equation}
J=\frac{E_{\rm APMA}-E_{\rm PMA}}{2S},
\end{equation}
where $E_{\rm APMA}$ and $E_{\rm PMA}$ are the minimum total energy for APMA and PMA, respectively, and $S$ is the interface area.
For results after optimization, we divide the numerical difference by the interface area of the system with stable magnetization alignment
(the results of the change rate of interface area between PMA and APMA are less than 1\% in all systems).
The magnetization shown in TABLE \ref{table:opt} is defined as the sum of local magnetic moments for stable magnetization alignment per unit volume.
In each system, the total electronic energy after optimization is less than that before optimization by approximately $5 \sim 10 $ eV.
The values of magnetization after optimization are close to those before optimization,
while the values of coupling energy after magnetization are not so close to those before magnetization (but the signs of coupling constants are consistent).
The coupling constants of NFB(001)/Fe(100) and NFB(100)/Fe(110) before optimization
are consistent with the previous results \cite{TogaMoriyaTsuchiuraSakuma2011}.
Among all our models, anti-parallel states are stable only in NFB(100)/Fe(110).
The positive coupling constant of NFB(001)/Fe(100) and
the negative coupling constant of NFB(100)/Fe(110) are in good agreements with
the recent experimental results of Ogawa et al. \cite{Ogawa2015}.
Our results for NFB(110)/Fe(112) indicate positive coupling constant,
whereas a negative coupling constant for NFB(110)/bcc-Fe
has also been reported by Ogawa et al.
However, it should be noted that the crystallographic alignment of the bcc-Fe layer has not yet been specified in that study.

\begin{table}[htbp]
\centering
\caption{\raggedright
Magnetic properties before and after optimization.
$J$ and $M$ are the results of coupling constant and magnetization after optimization
with respect to the cell volume, cell shape, and ion sites, respectively.
The subscripts $0$ represent the results before optimization.}
\label{table:opt}
\begin{tabular}{c|c c c c}
\hline
\hline 
 & $J_0$ {[}J/m$^{2}${]} & $J$ {[}J/m$^{2}${]} & $M_0$ {[}T{]} & $M$ {[}T{]} \tabularnewline
\hline
NFB single phase  & - & - & - & 1.57 \tabularnewline
NFB(001)/Fe(100) & 0.060 & 0.044 & 1.62 & 1.65 \tabularnewline
NFB(100)/Fe(110) & -0.013 & -0.023 & 0.49 & 0.58 \tabularnewline
NFB(110)/Fe(112) & 0.005 & 0.011 & 1.72 & 1.72 \tabularnewline
NFB(001)/Co(0001) & 0.030 & 0.055 & 1.43 & 1.46 \tabularnewline
NFB(100)/Co(0001) & 0.10 & 0.12 & 1.57 & 1.61 \tabularnewline
NFB(110)/Co(0001) & 0.11 & 0.16 & 1.57 & 1.56 \tabularnewline
\hline 
\hline
\end{tabular}
\end{table}

The comparison between the results of the TM = Fe system and those of the TM = Co system show that 
the optimized interlayer distance of the former models are shorter than those of the latter due to the strength difference of the nuclear interactions.
The absolute value of coupling energy for TM = Fe is smaller than that for TM = Co,
which may reflect the situation that the magnetic stiffness of Fe is lower than that of Co whose $d$-electron number is optimum for ferromagnetism.
It is confirmed from TABLE \ref{table:opt} that the
magnetization of NFB(001)/Fe(100), NFB(110)/Fe(112), and NFB(100)/Co(0001)
are stronger than that of the NFB single phase,
and hence, these remanent magnetizations are expected to be improved in real materials.
In particular, the magnetizations of TM = Fe systems are higher than those of TM = Co systems
because of the larger local moment of TM = Fe systems.
As the degree of magnetization depends on the volume of the unit cell,
a higher magnetization is exhibited for a shorter interlayer distance.

\begin{figure}[htbp]
	\begin{minipage}[b]{0.49\columnwidth}
		\centering
		\subcaption{TM = Fe(110)}
		\includegraphics[width=1.0\columnwidth]{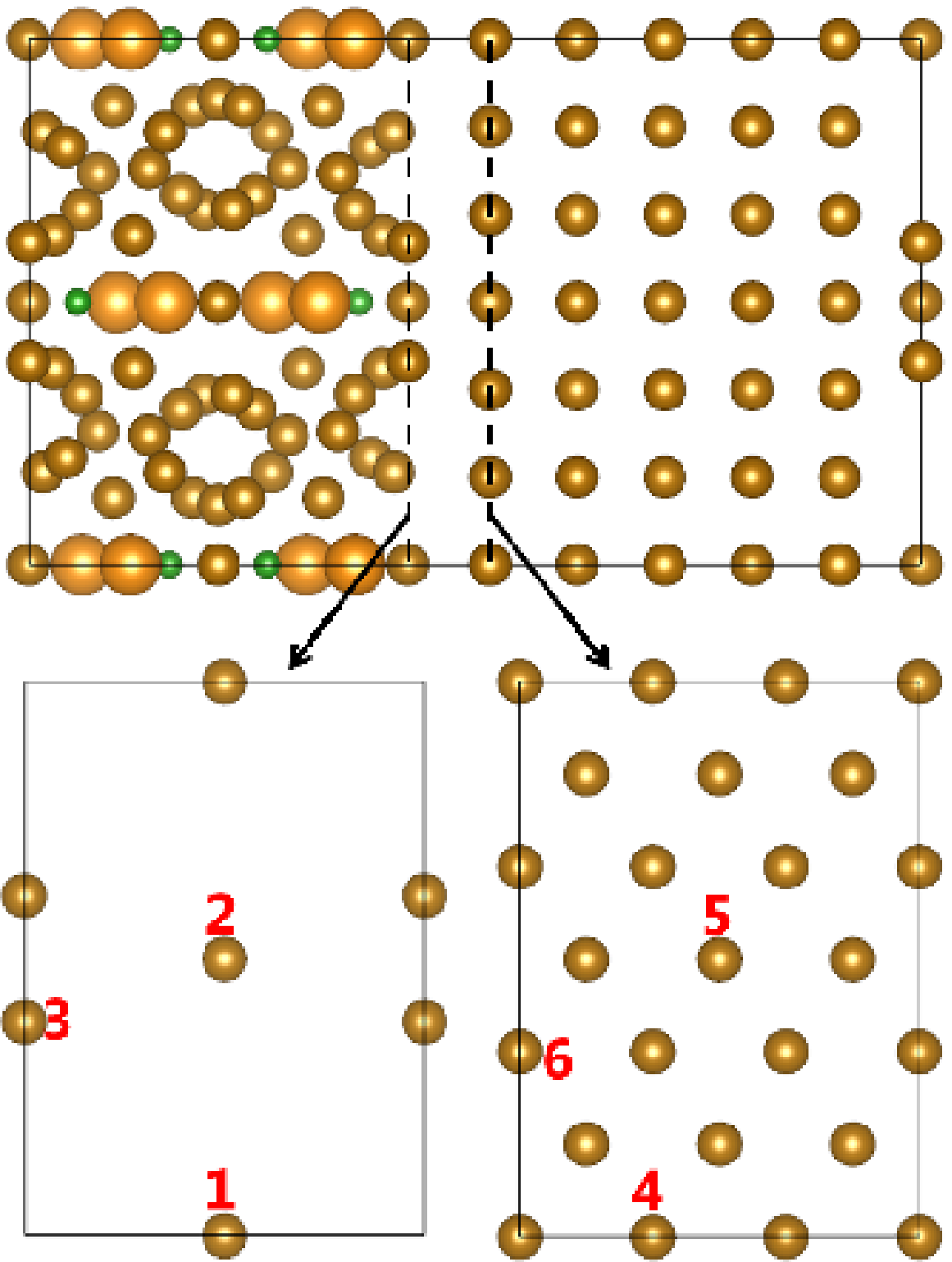}
		\label{fig:suf_fe}
	\end{minipage}
	\begin{minipage}[b]{0.49\columnwidth}
		\centering
		\subcaption{TM = Co(0001)}
		\includegraphics[width=1.0\columnwidth]{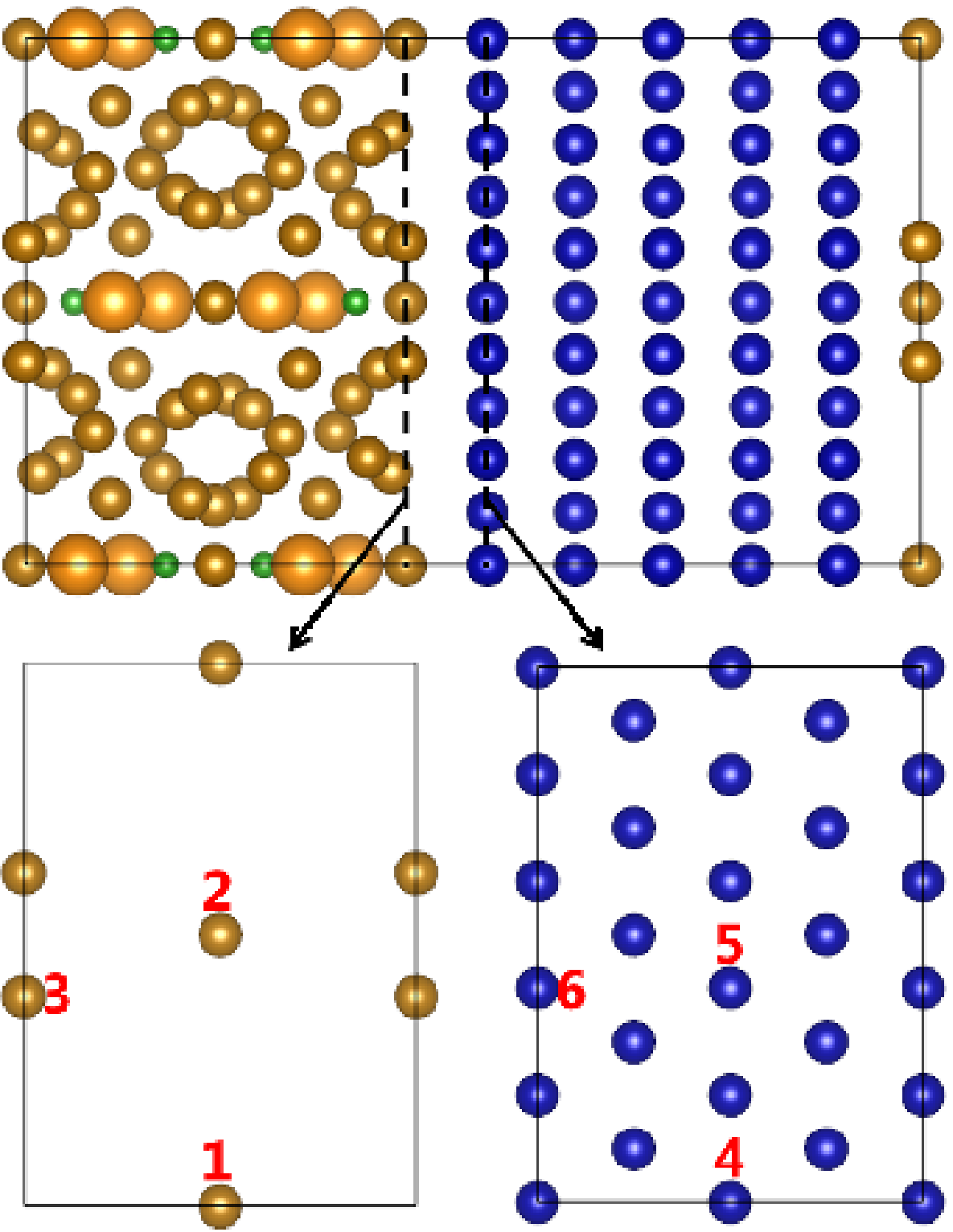}
		\label{fig:suf_co}
	\end{minipage}
	\caption{\raggedright
	Ion configurations at the NFB(100)/TM interface.
	The local magnetic moments of numbered ions are discussed in main text.}
	\label{fig:surface}
\end{figure}

To understand the unique property of NFB(100)/Fe(110),
we analyze the local magnetic moments around the NFB(100)/TM interface.
The ion positions for which the local moments are discussed here are shown in FIG. \ref{fig:surface}
(numbered sites ``1,'' ``2,'' and ``3'').
There are four Fe ions on the surface of NFB(100) in the unit cell,
but we show the results of three of them 
because two lined up on the $c$-axis are at the symmetrical positions.
In addition to the local moments at these sites,
we also show three local moments on the surface of TM layers (numbered sites ``4,'' ``5,'' and ``6'' in FIG.\ref{fig:surface}),
which are the nearest neighbors to sites ``1,'' ``2,'' and ``3,'' respectively.

\begin{figure}[htbp]
\centering
	\begin{minipage}[b]{0.494\columnwidth}
		\centering
		\includegraphics[width=1.0\columnwidth]{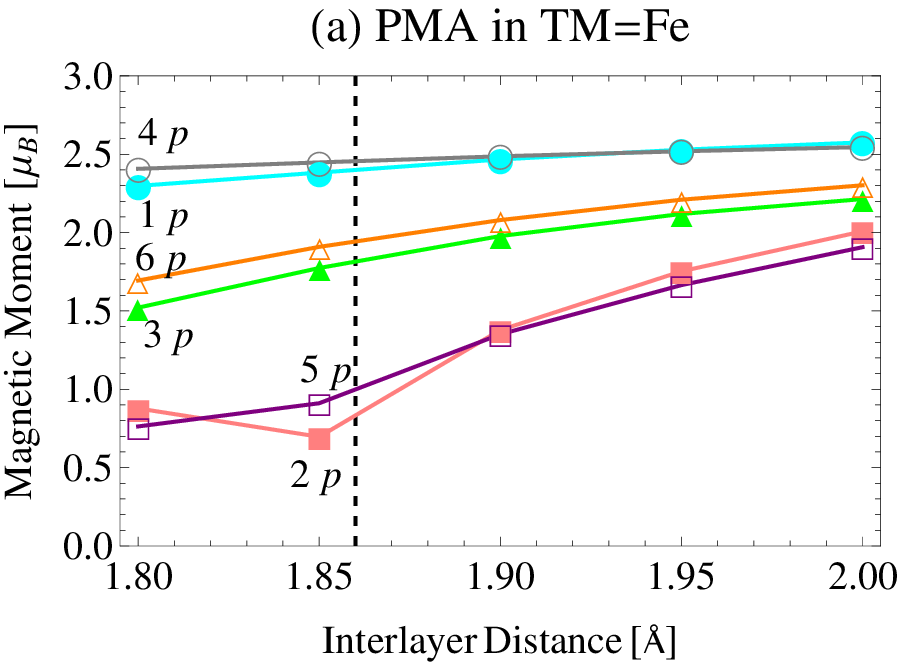}
	\end{minipage}
	\begin{minipage}[b]{0.494\columnwidth}
		\centering
		\includegraphics[width=1.0\columnwidth]{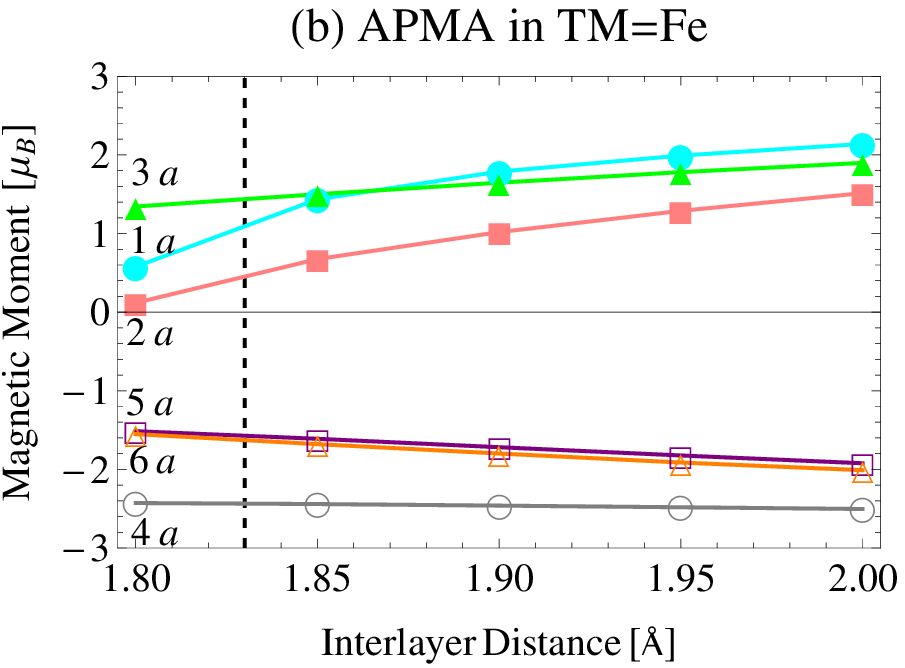}
	\end{minipage}
	\begin{minipage}[b]{0.494\columnwidth}
		\centering
		\includegraphics[width=1.0\columnwidth]{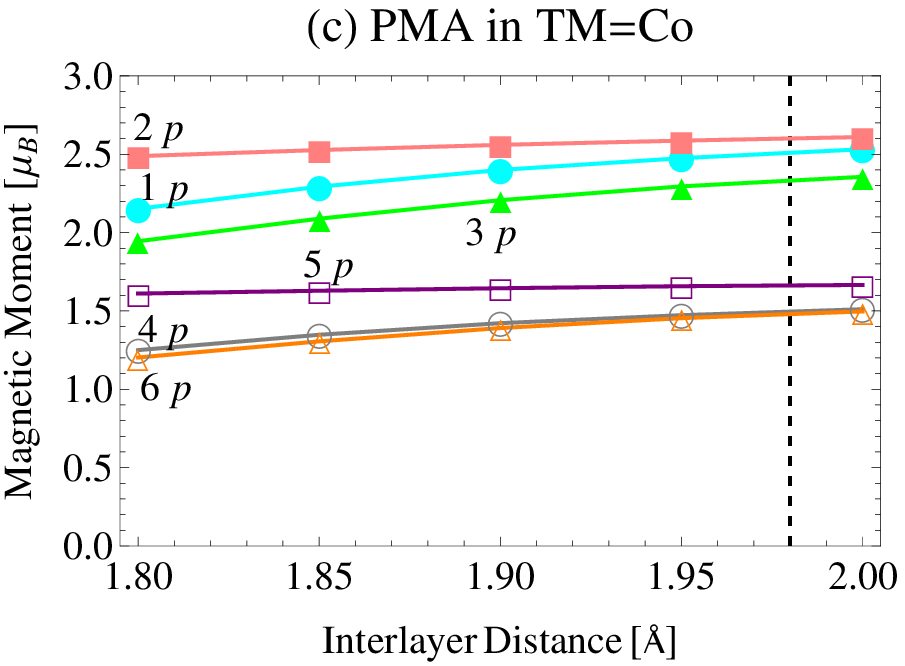}
	\end{minipage}
	\begin{minipage}[b]{0.494\columnwidth}
		\centering
		\includegraphics[width=1.0\columnwidth]{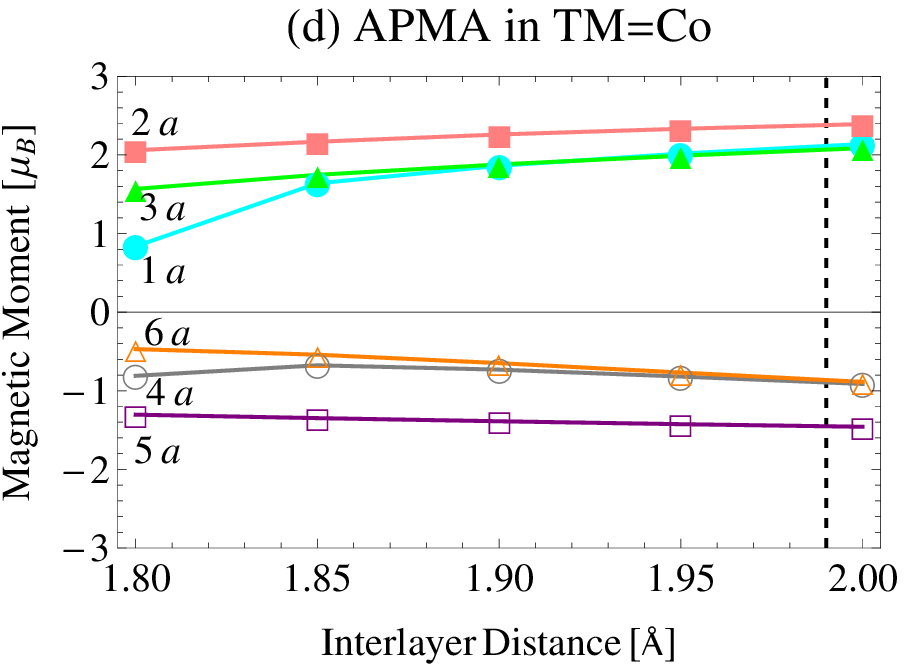}
	\end{minipage}
	\caption{\raggedright
	Magnetic moments at the numbered sites of the NFB(100)/TM interface shown in FIG. \ref{fig:surface}.
	The dashed lines represent the optimized interlayer distance for each system.}
	\label{fig:locmag}
\end{figure}

The interlayer-distance dependence of the local magnetic moments at each site are shown in Fig. \ref{fig:locmag}
[FIG. \ref{fig:locmag} (a) and (b) show the results of the TM = Fe system for PMA and APMA, respectively,
and FIG. \ref{fig:locmag} (c) and (d) show the results of the TM = Co system for PMA and APMA, respectively].
We deliberately show the non-optimized results here in order to simplify the discussion.
The absolute values of the local moment increase with increasing interlayer distance, $d$,
owing to the electron localization, and these values approach the saturation values.
Since the saturation magnetization of Co is less than that of Fe, 
the absolute values of $M_4$, $M_5$, and $M_6$ (the magnetic moment at site ``$n$'' is represented by $M_n$) of the TM = Co system
are smaller than those of the TM = Fe system.

It is confirmed from FIG. \ref{fig:locmag} (a) that
the nearest-neighbor local moments are strongly coupled to each other
because the behaviors of $M_{1p}$, $M_{2p}$, and $M_{3p}$ (the subscript ``$p$'' indicates PMA)
are almost similar to those of $M_{4p}$, $M_{5p}$, and $M_{6p}$, respectively.
These values reflect the strength of electron localization, which depends on inter-site distances
(e.g., when $d=1.8$, the inter-site distance between sites ``1'' and ``4,'' $l^{\rm Fe}_{1-4}$, is 2.32 \AA,
$l^{\rm Fe}_{2-5}$ is 1.8 \AA, and  $l^{\rm Fe}_{3-6}$ is 1.91 \AA).
The decreasing behavior of $M_{2p}$ is observed from 1.8 \AA\, to 1.85 \AA\, unlike the other local moments.
We calculate $M_{2p}$ in more detail to be less than $d = 1.90$ \AA,
and find that $M_{2p}$ shows variational behavior with a small change in $d$ below $d = 1.83$ \AA.
It is likely that too small a value of $d(=l^{\rm Fe}_{2-5})$ would result in overlapping atoms
and the local moment being not appropriately evaluated.
However, these unexpected results are not of major importance to our qualitative discussion.

\begin{figure}[htbp]
	\begin{minipage}[b]{0.494\columnwidth}
		\centering
		\includegraphics[width=1.0\columnwidth]{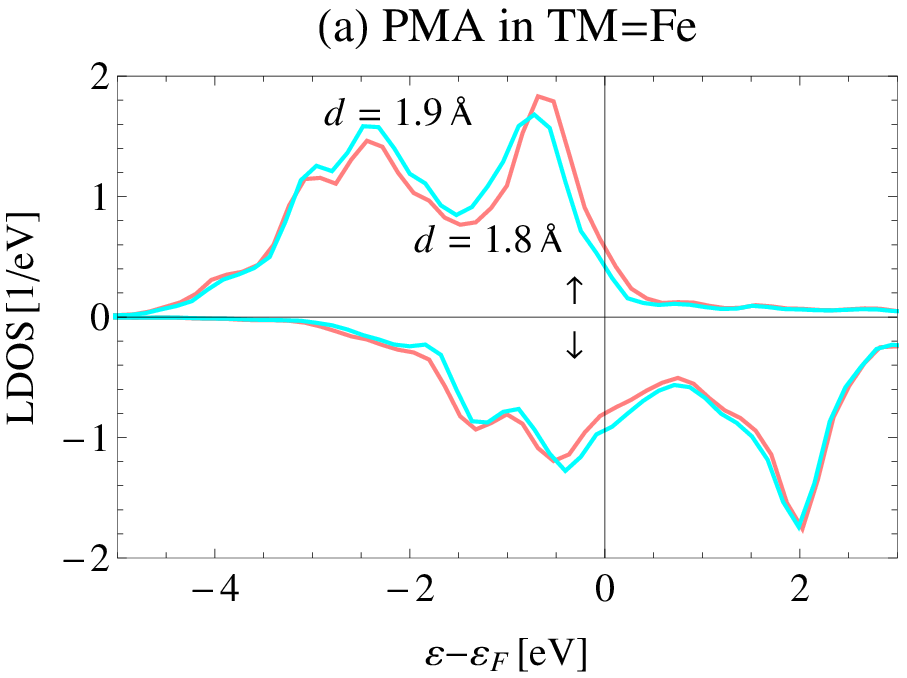}
	\end{minipage}
	\begin{minipage}[b]{0.494\columnwidth}
		\centering
		\includegraphics[width=1.0\columnwidth]{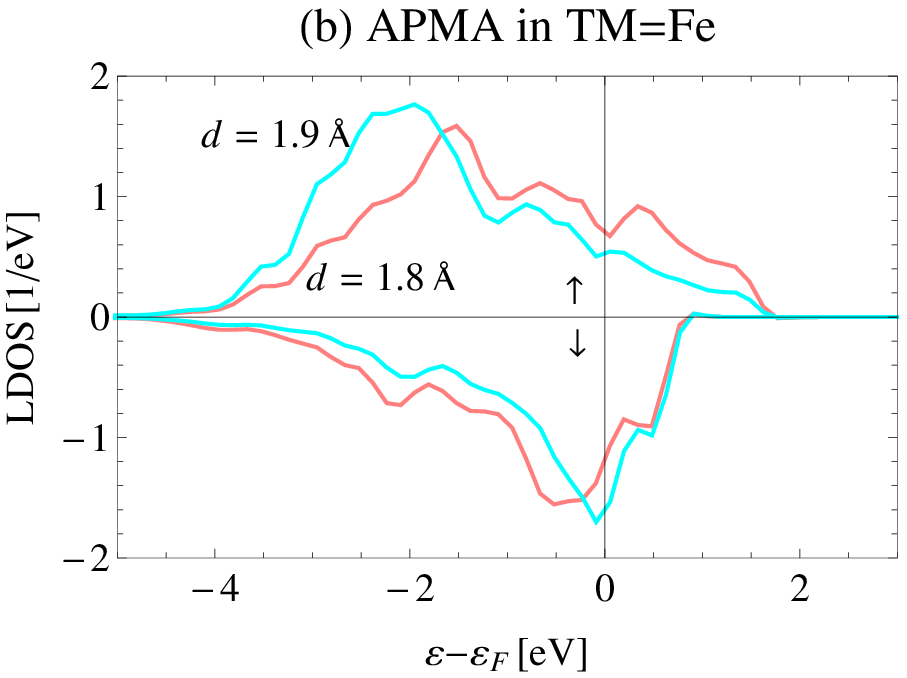}
	\end{minipage}
	\begin{minipage}[b]{0.494\columnwidth}
		\centering
		\includegraphics[width=1.0\columnwidth]{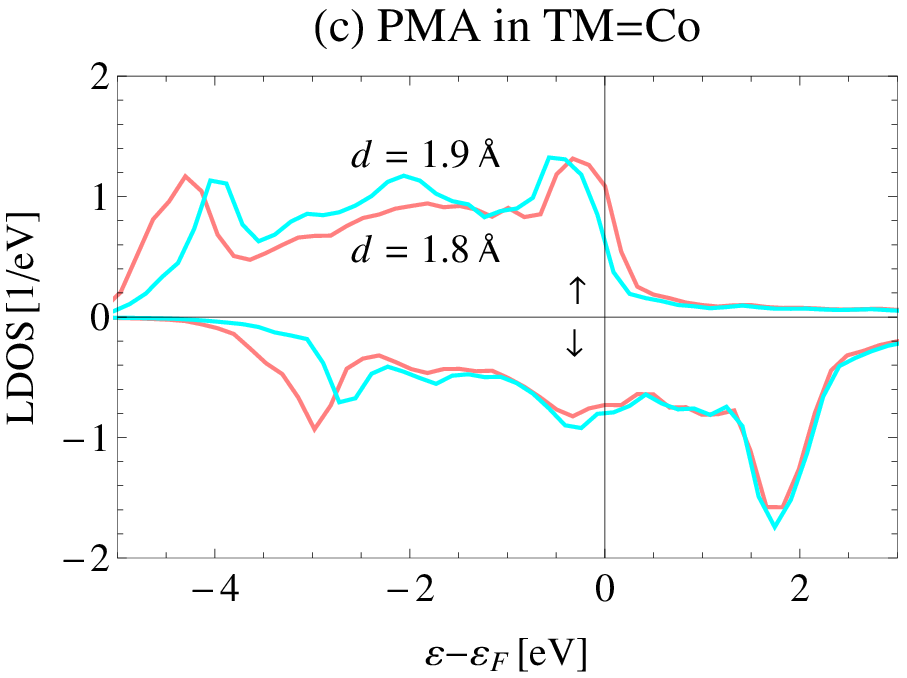}
	\end{minipage}
	\begin{minipage}[b]{0.494\columnwidth}
		\centering
		\includegraphics[width=1.0\columnwidth]{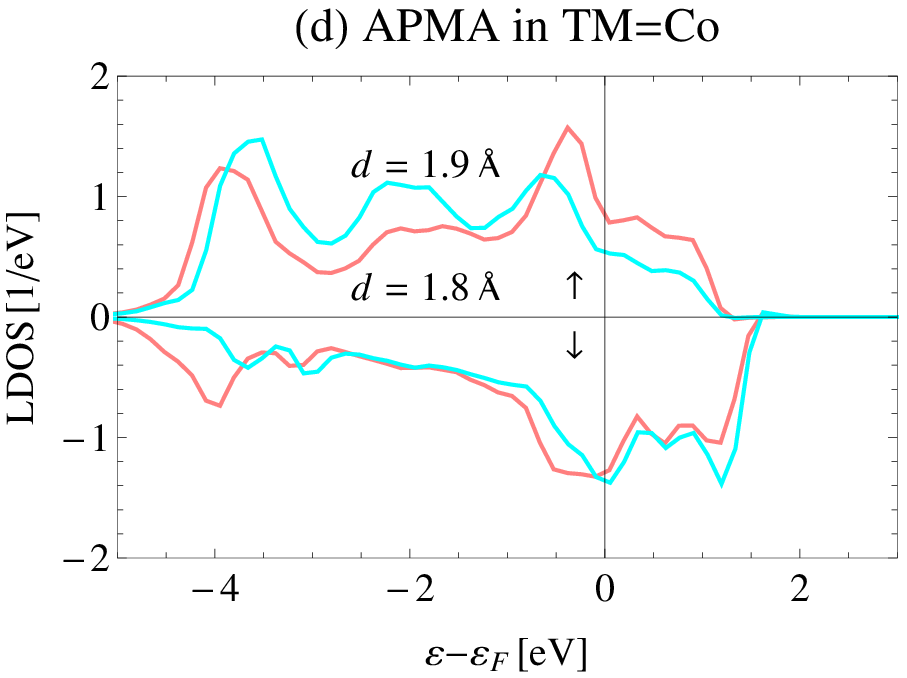}
	\end{minipage}
	\caption{\raggedright
	Local density of states (LDOS) of $d$-electrons at site ``1'' in NFB(100)/TM.
	The origin of the horizontal axis is the Fermi energy.
	\label{fig:ldos}}
\end{figure}

In FIG. \ref{fig:locmag} (b), a clear increase of $M_{1a}$ (the subscript ``$a$''  indicates APMA) is observed from 1.8 \AA\ to 1.9 \AA,
while $M_{4a}$, the site of which is the nearest neighbor to site ``1,'' hardly depends on $d$.
These results indicate that $M_1$ is easily coupled to the magnetic moments of the TM layer, rather than to those of NFB.
We confirm that the local density of states (LDOS)  at site ``1'' for APMA strongly depend on $d$,
whereas the LDOS at site ``1'' for PMA hardly depend on it
[see FIG. \ref{fig:ldos} (a) and (b)].
Figure \ref{fig:ldos} (b) shows that the band splitting is reduced with decreasing $d$,
which implies an enhancement of the electron itinerant property.
A reduction in band splitting with decreasing $d$ is also observed at site ``2'' for APMA,
while it is not clearly observed in the other systems.
We conclude that itinerant magnetism at the interface is preferred in APMA more than in PMA
and that it lowers the total electron energy by overcoming the increase of exchange coupling energy.

The local moments for PMA in the TM = Co system shown in FIG. \ref{fig:locmag} (c)
reflects the coupling strength, which depends on inter-site distances, similar to FIG. \ref{fig:locmag} (a)
(e.g., when $d=1.8$, the inter-site distance between sites ``1'' and ``4,'' $l^{\rm Co}_{1-4}$, is 1.8 \AA,
$l^{\rm Co}_{2-5}$ is 2.17 \AA, and  $l^{\rm Co}_{3-6}$ is 1.81 \AA).
These local moments shown in FIG. \ref{fig:locmag} (c) are not as dependent on $d$ as those in FIG. \ref{fig:locmag} (a) are,
because the magnetic stiffness of Co is greater than that of Fe and the magnetization of Co is less than that of Fe.
$M_a$ in the TM = Co system clearly increases from 1.8 \AA\ to 1.9 \AA,
and it is confirmed from FIG. \ref{fig:ldos} (c) and (d) that
the band splitting at site ``1'' for APMA is reduced with decreasing $d$ by more than that for PMA is.
These results are similar to the results for the TM = Fe system,
while the behavior of $M_{2a}$ and LDOS at site ``2'' are different between the TM = Co and TM = Fe systems.
The inter-site distance between site ``2'' and ``5'' of the TM = Co system is greater than that of the TM = Fe system
(e.g., when  $d=1.8$, $l^{\rm Co}_{2-5}$ is greater than than $l^{\rm Fe}_{2-5}$ by 0.37 \AA)
and a reduction in band splitting with increasing $d$ is not clearly observed at site ``2'' in the TM = Co system, unlike in the TM = Fe system.
This band-splitting effect is not observed in other systems of TM = Co,
and hence, at the interface of the TM = Co system, itinerant magnetism does not grow as much as in the TM = Fe system.
Moreover, a noticeable decrease in $M_{1a}$, which reflects an enhancement of the itinerant property,
occurs near the optimized value of $d$ in the case of TM = Fe,
whereas in the case of TM = Co, a noticeable decrease occurs at a value of $d$ much smaller than the optimized value.
Therefore, anti-parallel states are less likely to occur in real materials of TM = Co than in those of TM = Fe.
The same can be said of NFB(001)/TM and NFB(110)/TM
because their optimized interlayer distances are much greater than those of NFB(100)/TM (see FIG. \ref{fig:freeng}).

It should be noted that the values of $M_{3a}$ in both the TM = Fe and Co systems are not as dependent on $d$ as the values of $M_{1a}$ are in the TM = Fe and Co systems,
even though $l^{\rm Fe}_{3-6}$ and $l^{\rm Co}_{3-6}$ are not as long as $l^{\rm Fe}_{1-4}$ and $l^{\rm Co}_{1-4}$.
One might think that this is attributed to the fact that $M_1$ is more isolated than $M_3$ owing to Nd ions
[i.e., site ``1'' is located in the (001) plane of NFB,
in which Nd ions are located, while site ``3'' is not located in this plane].
In practice, however, Nd ions do not affect the magnetic property of NFB(001)/Fe(110).
We calculate the NFB(100)/Fe(110) system by replacing Nd ions, which are the nearest neighbors to sites ``1'' and ``3'', with Fe ions,
and confirm that the magnetic properties and interface local moments of this substituted system are similar to those of the unsubstituted system.
Therefore, we conclude that the interface coupling property is mainly dependent on
the interface structure (i.e., inter-atomic distance, coordination number, and atomic positional relationship). 

\section{Summary}
We studied the interface magnetic properties of Nd${}_2$Fe${}_{14}$B (NFB)/TM (TM = Fe, Co) multilayer exchange spring magnets
on the basis of first-principles calculations.
Assuming a collinear spin structure, we optimized the structure of NFB(001)/TM, NFB(100)/TM, and
NFB(110)/TM, and discussed their resultant magnetic properties.
Improvements in the remanent magnetization were observed in NFB(001)/Fe(100), NFB(110)/Fe(112), and NFB(100)/Co(0001),
and it was found, as expected, that the model with TM = Fe, rather than that with TM = Co, is advantageous to remanence.
The optimized interlayer distance is shortest for NFB(100)/Fe(110), 
and anti-parallel magnetization alignment between NFB and Fe layers is preferred.
From the analysis of interface magnetic moments and local DOS of NFB(100)/Fe(110), it was found that interface exchange splitting is reduced with decreasing interlayer distance.
Therefore, the total energy of NFB(100)/Fe(110) is lowered by enhancing the interface itinerant property through anti-ferromagnetic coupling.
On the other hand, in NFB(100)/Co(0001),
interface magnetic moments are not so dependent on the interlayer distance, because the magnetic stiffness of Co is greater than that of Fe.
From the analysis of local DOS,
it was confirmed that the interface itinerant property of NFB(100)/Co(0001) is lower than that of NFB(100)/Fe(110)
owing to the difference of interface structures.
Moreover, the optimized interlayer distance of NFB(100)/Co(0001) is greater than that of NFB(100)/Fe(110),
and it was predicted that interface anti-ferromagnetic coupling is less likely to occur in real materials of NFB(100)/Co(0001).
Thus, it is concluded that the magnetic performance of exchange spring multilayer magnets is strongly related to
inter-site distance around the interface and magnetic stiffness of the soft magnetic phase.
Proper material selection for the soft magnetic layer and more precise surface control techniques
are indispensable for improving the properties of exchange spring magnets.



%
%

%

\begin{acknowledgments}
This work was supported by CREST-JST.
\end{acknowledgments}


%

\end{document}